\documentclass[showpacs,prl,twocolumn,superscriptaddress,notitlepage,nofootinbib]{revtex4-2}
\usepackage[dvips]{graphicx}
\usepackage{amsmath,amssymb,amsthm,mathrsfs,amsfonts,dsfont,hyperref,subfigure, epsfig, braket,bm,enumerate,color,graphicx,dcolumn}
\usepackage[normalem]{ulem}
\usepackage[dvipsnames]{xcolor}
\usepackage{{xcolor,charter, gensymb, float}}
\newcommand{\ctc}{Cs$_2$TaCl$_6$}
\usepackage [english]{babel}
\begin{document}
\sloppy

\urldef{\github}\url{https://github}

\title{Charge multipoles correlations and order in Cs$_2$TaCl$_6$}

\author{Aria Mansouri Tehrani$^*$}
\affiliation{Materials Theory, ETH Zurich, Wolfgang-Pauli-Strasse 27, 8093 Zürich, Switzerland}
\author{Jian-Rui Soh$^*$}
\affiliation{Institute of Physics, Ecole Polytechnique Federale de Lausanne (EPFL), CH-1015 Lausanne, Switzerland}
\author{Jana Pásztorová}
\affiliation{Institute of Physics, Ecole Polytechnique Federale de Lausanne (EPFL), CH-1015 Lausanne, Switzerland}
\author{Maximilian E. Merkel}
\affiliation{Materials Theory, ETH Zurich, Wolfgang-Pauli-Strasse 27, 8093 Zürich, Switzerland}
\author{Ivica Živković}
\affiliation{Institute of Physics, Ecole Polytechnique Federale de Lausanne (EPFL), CH-1015 Lausanne, Switzerland}
\author{Henrik M. Rønnow}
\affiliation{Institute of Physics, Ecole Polytechnique Federale de Lausanne (EPFL), CH-1015 Lausanne, Switzerland}
\author{Nicola A. Spaldin}
\affiliation{Materials Theory, ETH Zurich, Wolfgang-Pauli-Strasse 27, 8093 Zürich, Switzerland}
\email{nicola.spaldin@mat.ethz.ch}
\def\thefootnote{*}\footnotetext{Aria Mansouri Tehrani and Jian-Rui Soh contributed equally to this work}

\begin{abstract}

We examine the role of charge, structural, and spin degrees of freedom in the previously poorly understood phase transition in the 5$d^1$ transition-metal double perovskite \ctc\, using a combination of computational and experimental techniques. Our heat capacity measurements of single-crystalline \ctc\, reveal a clear anomaly at the transition temperature, $T_\mathrm{Q}$, which was not previously observed in polycrystalline samples. Density functional calculations indicate the emergence of local charge quadrupoles in the cubic phase, mediated by the paramagnetic spins or local structural distortions which then develop into long-range ordered charge quadrupoles in the tetragonal phase. Our resonant elastic x-ray scattering on \ctc\, single crystals lend support to our calculations. Our work provides new insight into the phase transition in \ctc\, at $T_\mathrm{Q}$, and demonstrates the utility
of this combination of techniques in understanding the complex physics of hidden orders in paramagnetic spin-orbit-entangled compounds.

\end{abstract}
\maketitle

In $3d^1$ transition-metal perovskites, the on-site Hubbard Coulomb repulsion, one-electron bandwidth, magnetic exchange, crystal-field splitting and lattice strain energies are of similar size, and their competition results in a rich variety of ground states, ranging from correlated metals (SrVO$_3$ and CaVO$_3$)\cite{morikawa_spectral_1995} to antiferromagnetic (LaTiO$_3$)\cite{hays_electronic_1999} and ferromagnetic (YTiO$_3$)\cite{zhou_evidence_2005} insulators. The crystal structures exhibit symmetry-lowering phase transitions to states with tilts and rotations of the oxygen octahedra, as well as Jahn-Teller distortions related to the lifting of orbital degeneracy. 


In $5d^1$ {\it double} perovskites, in which half of the B-site cations are either absent or replaced by non-magnetic ions, the B-site dilution reduces the bandwidth so that it is again comparable in energy to the other interactions listed above; enhanced spin-orbit coupling from the heavy B-site cations and magnetic frustration associated with their face-centered cubic arrangement further increase the manifold of competing interactions. The phase diagram resulting from a model Hamiltonian containing these interactions \cite{chen_exotic_2010} includes ferromagnetic and antiferromagnetic phases with magnetic octupolar order dominating over the usual dipolar order, and charge quadrupole-ordered paramagnetism, as well as a possible quantum-spin-liquid state. 

Experimental studies of $5d^1$ double perovskites have focused on the oxides such as Ba$_2$NaOsO$_6$ and Ba$_2$MgReO$_6$, both of which show two phase transitions in measurements of their specific heat as a function of temperature. The low-temperature phase transition, $T_\mathrm{M}$, occurring at $~\sim{7}$\,K and $~\sim{18}$\,K respectively, corresponds to the formation of the magnetically ordered state, which in both cases is a strongly canted antiferromagnet with a net magnetization \cite{liu_phase_2018, lu_magnetism_2017, hirai_successive_2019, hirai_detection_2020}.
The higher-temperature transition (at $T_\mathrm{Q}$ $~\sim{10}$\,K for Ba$_2$NaOsO$_6$ and $~\sim{33}$\,K for Ba$_2$MgReO$_6$) has been attributed to the ordering of charge quadrupoles on the $d^1$ B-site cations  and is accompanied by a small structural distortion, detectable only with high-resolution x-ray diffraction\cite{liu_phase_2018, lu_magnetism_2017, hirai_successive_2019, hirai_detection_2020, hirai_possible_2021}. Recently, the chlorides $A_2$TaCl$_6$ ($A$ = K, Rb, Cs) have also attracted some attention, motivated by their greater ionicity which positions them closer to the formal $d^1$ limit\cite{ishikawa_ordering_2019,yun_dicaesium_2007}. In this work, we focus our attention on \ctc.

\ctc\, is a vacancy-ordered double perovskite which crystallizes in the cubic F$m\bar{3}m$ space group at room temperature with lattice constant $a$ = 10.3285(4)\,\AA\, based on experimental powder x-ray diffraction data\cite{ishikawa_ordering_2019}. Heat capacity measurements indicate an almost complete recovery of the electronic entropy of Rln4, which corresponds to the ideal J$_{eff}$=3/2 quartet state as expected for a $5d^1$ Ta ion in an ideal octahedral environment. Upon cooling, the Rln4 can be accounted for by assuming, first, the ordering of the multipolar (magnetic or charge) degree of freedom, followed by the ordering of magnetic dipoles. Whilst the latter is seen from the pronounced peak in the heat capacity measurements at $T_\mathrm{M}\sim4$\,K, no clear signature of the former has been seen in earlier studies on the powder sample of \ctc. Instead, it has been proposed that in addition to the ordering of magnetic dipoles, fluctuating short-ranged orders of entangled multipoles such as charge quadrupoles are responsible for the full recovery of $Rln4$\cite{ishikawa_ordering_2019}. Furthermore, it has been shown that \ctc\, undergoes a structural phase transition to tetragonal symmetry ($a$ = 7.29\,\AA, and $c$ = 10.16 lattice parameters) with I4/$mmm$ space group around 30\,K; this symmetry allows a ferroic ordering of the charge quadrupoles on the Ta sites\cite{ishikawa_ordering_2019}.

In the current manuscript, we combine state-of-the-art computational and experimental techniques to study the previously poorly understood phase transition reported in \ctc\, at $T_\mathrm{Q}$. Specifically, we revisit the specific heat measurements, where surprisingly no anomaly was previously observed\cite{ishikawa_ordering_2019} and explore the nature of the proposed multipolar fluctuations around the phase transition\cite{ishikawa_ordering_2019}. To achieve these objectives, we synthesize polycrystalline and single-crystalline \ctc\, and perform a comparative study of their heat capacities. Second, we use single-crystal x-ray diffraction to detect structural changes at $T_\mathrm{Q}$. In addition, we use density functional theory (DFT) and resonant x-ray diffraction to look for evidence of local multipoles and their entanglements to the structural phase transition in the vicinity of $T_\mathrm{Q}$. Finally, we discuss the possible role of higher-order charge multipoles at the transition at $T_\mathrm{Q}$.

\begin{figure}[t!]
\includegraphics[width=3.3in]{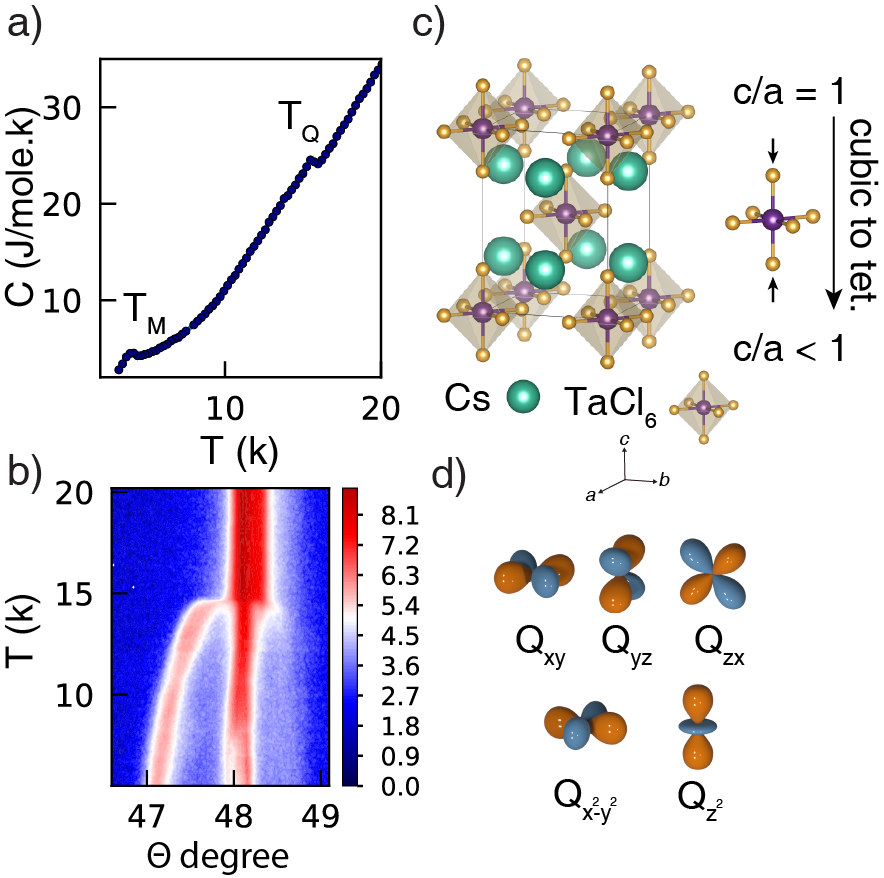}
\centering
\caption{a) Measured heat capacity of single-crystalline \ctc\, as a function of temperature. The heat capacity curves display an anomaly at $T_\mathrm{M}$, and an additional kink at $T_\mathrm{Q}$. b) Synchrotron x-ray diffraction pattern of \ctc\, as a function of $T$ shows a splitting of the (777)$_\mathrm{c}$ peak below $T_\mathrm{Q}$. c) Crystal structure of \ctc\, in the reduced unitcell representation. The room temperature cubic unit cell (with cell parameter, $a_c$) can also be depicted in the tetragonal setting by adopting the tetragonal lattice parameters $c_t$=$\sqrt{2}a_t$=$a_c$, where the TaCl$_6$ octahedra are undistorted. Here, the subscripts $c$ and $t$ refer to the cell parameters (and ($hkl$) Miller indices) in the cubic and tetragonal setting, respectively. d) Sketch of the charge quadrupoles in cartesian representation. The orange and cyan regions represent regions with excess and reduced electronic charge.}
\label{Fig1}
\end{figure}


We first present our heat capacity measurements and analyze them for indications of phase transitions, particularly of multipolar order. Figure~\ref{Fig1}a shows the heat capacity of our single-crystalline \ctc\, as a function of temperature. An anomaly at $T_\mathrm{M}$=4.6\,K is observed in the heat capacity curve, which is concomitant with the onset of the magnetic dipolar order of Ta ions and is consistent with an earlier report~\cite{ishikawa_ordering_2019} and our measurements of the heat capacity of polycrystalline \ctc\, as shown in our supporting information. Furthermore, the temperature dependence of the heat capacity of single-crystalline \ctc\, also displays a kink at $T_\mathrm{Q}\sim$15\,K, which was not observed in our polycrystalline samples or reported in the polycrystalline samples of Ref.~\onlinecite{ishikawa_ordering_2019}. The observation of a peak in the heat capacity at $T_\mathrm{Q}$ in single-crystalline \ctc\, clearly indicates a phase transition at this temperature. This resolves the earlier apparent ``mystery" regarding the absence of an anomaly in the heat capacity in polycrystalline \ctc\, which was attributed to the build-up of a hidden order [\onlinecite{ishikawa_ordering_2019}]. One possible reason why polycrystalline \ctc\, (both in Ref.\onlinecite{ishikawa_ordering_2019} and our work) does not manifest an anomaly at $T_\mathrm{Q}$ could be the small size of the polycrystalline \ctc\, grains broadening the transition. 

Having unambiguously established the presence of a phase transition at $T_\mathrm{Q}$, we next study the relative roles of charge quadrupoles, magnetic dipoles, and lattice distortions in driving the transition, considering the temperature regimes $T<T_\mathrm{Q}$ and $T>T_\mathrm{Q}$ in turn.


Now we study the evolution of the crystal structure of single-crystalline \ctc\, as a function of temperature, using x-ray diffraction. The single-crystalline diffraction pattern of \ctc\, at $T=$300\,K is fully consistent with the $Fm\bar{3}m$ ($\#225$) space group, in good agreement with earlier reports~\cite{yun_dicaesium_2007,ishikawa_ordering_2019}. The refinement of our temperature-dependent polycrystalline x-ray diffraction patterns (see supporting information) indicates that \ctc\, undergoes a symmetry lowering below $T\sim$ 28\,K, to the tetragonal $I4/mmm$ ($\#139$) space group, consistent with Ref.\onlinecite{ishikawa_ordering_2019-1} where an accompanying anomaly in the heat capacity measurements was also absent (See supporting information). Structurally, this symmetry lowering manifests as a compression of the TaCl$_6$ octahedron along $z$ with a reduction of the lattice parameter $c_t$ such that $c_t< \sqrt{2}a_t$.

In Fig.~\ref{Fig1}b we show the temperature dependence of the single crystal x-ray diffraction of \ctc. The $(777)_c$ reflection shown in Fig.~\ref{Fig1}b splits below 15\,K into the $(707)_t$ peaks, concomitant with the observed anomaly in single-crystal heat capacity measurements. We, therefore, assign this structural transition to $T_\mathrm{Q}$. Note that this occurs at a significantly lower temperature compared to the polycrystalline \ctc\, reported in this work and in Ref.~\onlinecite{ishikawa_ordering_2019}. The $(777)_c \to (707)_t$ peak splitting at $T_Q$, is consistent with the cubic-to-tetragonal distortion seen in polycrystalline \ctc, as shown in Fig.~\ref{Fig1}c, and is due to the three-fold freedom in the choice of the compression axis ($z$) in the tetragonal distortion of a cubic structure (either along $a$, $b$ or $c$), which gives rise to three structural domains and hence the split peaks. Since the multiplicity of peaks in single-crystal diffraction is always one, the reason for the peak splitting of $(707)_t$ below $T_\mathrm{Q}$ is subtly different from the reason for the peak splitting observed in the polycrystalline diffraction pattern of \ctc, which is due to the multiplicity (See supporting information). Nonetheless, the physical origin of the peak splitting is the same, driven by the compression of the TaCl$_6$ octahedron.

\begin{figure*}[t!]
\includegraphics[width=6in]{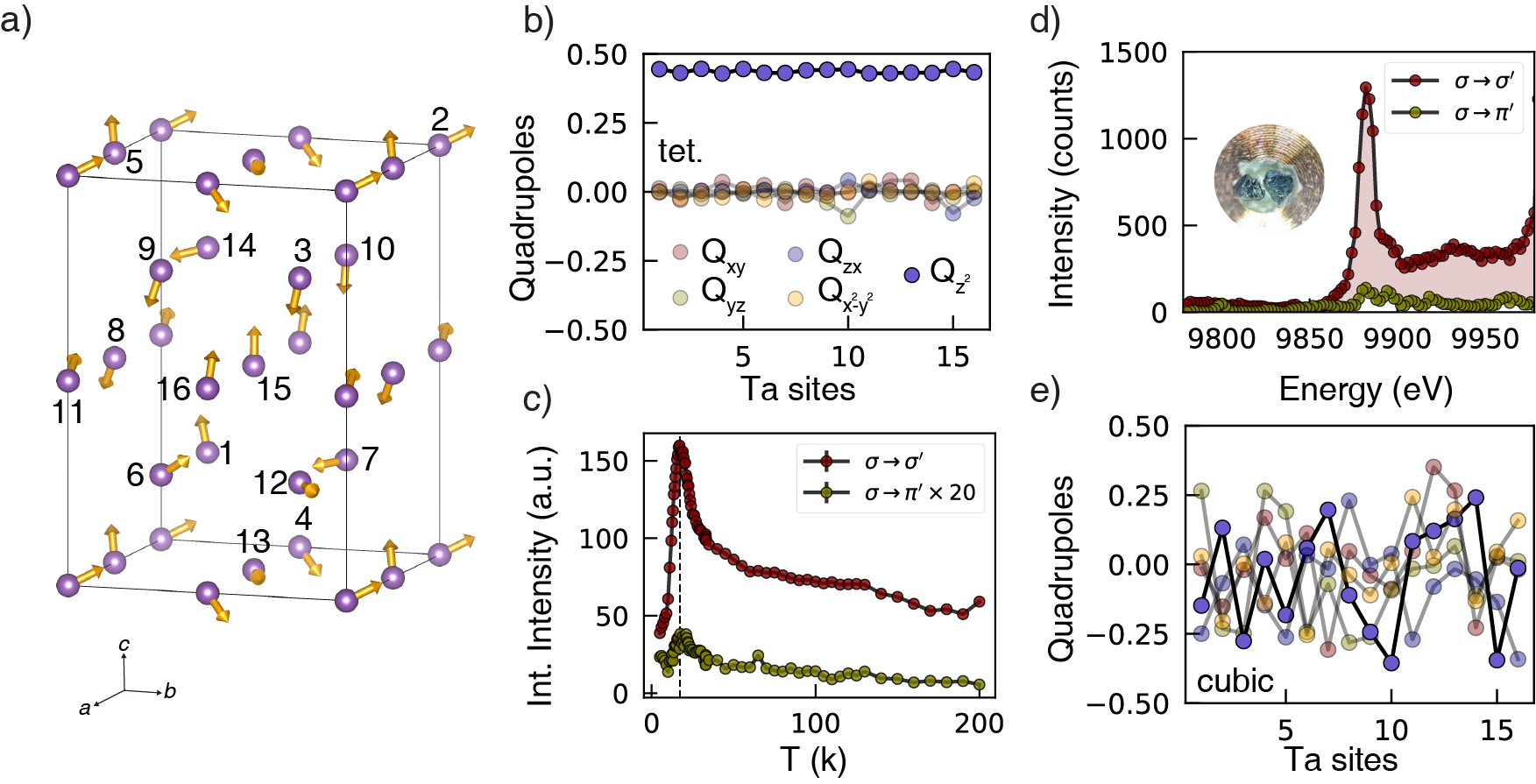}
\caption{a) Paramagnetic supercell of \ctc\, used for DFT calculations. Only Ta atoms are shown for clarity. b) Calculated charge quadrupoles on the Ta atoms labeled as in panel (a) in units of electron charges in the tetragonal crystal structure. c), Temperature dependence of the integrated intensity of the $(777)_\mathrm{c}$ and $(707)_\mathrm{t}$ reflections above and below $T_\mathrm{Q}$ (dashed line) measured with REXS. d) The summed intensity of the $(777)_\mathrm{c}$ reflection of \ctc\, as a function of incident x-ray energy measured at $T=20$\,K in the $\sigma$$\to$$\sigma^\prime$ and $\sigma$$\to$$\pi^\prime$ scattering channels. The insert shows the cleaved (111) faces of the flux-grown \ctc\, crystals. e) The magnitude of charge quadrupole components on the 16 Ta sites of the supercell of \ctc\, in units of electron charges in the cubic setting.}
\label{Fig2}
\end{figure*}

While the crystal structural distortion at $T_\mathrm{Q}$ as well as the anomaly in the heat capacity are indicative of the long-ranged ordering of charge quadrupoles, as pointed out previously\cite{ishikawa_ordering_2019}, they are not thoroughly studied yet. As the first step in this direction, we use first-principles calculations based on DFT to search for the various charge quadrupole components (Fig.~\ref{Fig1}b) and determine how they interact with the crystal structure. For \ctc, the B-site cations occupy centrosymmetric lattice sites, and so only even-ranked charge multipoles, charge quadrupoles and hexadecapoles, and odd-ranked magnetic multipoles, magnetic dipoles, and octupoles can be non-zero~\cite{suzuki_first-principles_2018}. In this work, we first focus on determining the existence and ordering of the charge quadrupoles and later discuss the possible role of the charge hexadecapoles in the phase transition at $T_\mathrm{Q}$.

Since standard density functional calculations with periodic boundary conditions yield the zero-kelvin magnetically ordered state, they are not immediately appropriate for describing the regime of interest here, in which the local magnetic dipoles on the Ta sites are not ordered. 
In this work, we adopt a method similar to that introduced in Ref.~\onlinecite{trimarchi_polymorphous_2018} and Ref.~\onlinecite{varignon_mott_2019}, which we recently applied to Ba$_2$MgReO$_6$\cite{mansouri_tehrani_untangling_2021}, and construct a supercell containing randomly oriented magnetic dipoles, constrained to have zero total magnetization, as shown in Fig.\,\ref{Fig2}c.

Fig.\,\ref{Fig2}b shows the calculated charge quadrupole components on the 16 Ta sites of our Cs$_2$TaCl$_6$ supercell constructed in the low-temperature experimental $I4/mmm$ crystal structure with the magnetic dipole moments constrained to random orientations with a net-zero total magnetization. The only non-zero charge quadrupoles in the paramagnetic state are the $Q_{z^2}$; these all have the same sign and are close to the same magnitude, indicating ferroic ordering. The ferroic ordering of the $Q_{z^2}$ quadrupoles is consistent with the crystal structure distortion that results in the contraction of Ta--Cl bonds along the $z$-axis by $\approx{0.4}$\,\AA.
The absence of any anti-ferroic ordering is consistent with the absence of other types of distortion, as also revealed by x-ray diffraction data. 

Having discussed the long-ranged order of the lattice distortion and charge quadrupoles in the paramagnetic phase below $T_\mathrm{Q}$, we now turn to the magnetic dipole, crystal structure, and charge degrees of freedom above $T_\mathrm{Q}$. In particular, we investigate the proposal of Ref.~\onlinecite{ishikawa_ordering_2019} proposed that short-ranged order in charge quadrupoles persists even when \ctc\, is in its high-symmetry cubic crystal structure.

We use resonant elastic x-ray scattering to look for evidence of short-ranged ferroic order of $Q_{z^2}$ charge quadrupoles above the structural transition. Performing REXS at the Ta $L_3$ edge has the three-fold benefit of (1) direct coupling to the Ta ions, (2) strong resonant enhancement of the scattered signal, and (3) sensitivity to any asphericity of the $5d^1$ electron cloud density which yields directly the charge quadrupoles.

In the $Fm\bar{3}m$ space group, our structure factor calculation indicates that the coherently scattered x-rays arising from the ferroic $Q_{z^2}$ charge quadrupole order coincides with the structural peaks in reciprocal space. Differentiating between the different contributions to the intensity of the scattered x-rays is difficult since both scattering processes benefit from resonant enhancement. To minimize the structural contributions to the scattered intensity, we chose a reflection with a small structure factor, namely the $(777)_c$ peak. Figure~\ref{Fig2}c shows the temperature dependence of the integrated intensity of the peak in the $\sigma$$\to$$\sigma^\prime$ and $\sigma$$\to$$\pi^\prime$ scattering channels. Notably, at all temperatures above $T_\mathrm{M}$, we observe that the intensity is mainly in the $\sigma$$\to$$\sigma^\prime$ channel, implying that the observed peak intensity is dominated by scattering processes that preserve the photon polarization. Such behavior arises from charge quadrupoles and the crystal structure but not from magnetic dipoles.

Our main result is an anomalous increase in the intensity of the $(777)_\mathrm{c}$ reflection in the $\sigma$$\to$$\sigma^\prime$ scattering channel on cooling towards $T_\mathrm{Q}$. Given that the structure of \ctc\, does not change above $T_\mathrm{Q}$, this additional scattering intensity cannot be explained by conventional Bragg scattering. In addition, we see a strong peak at the Ta $L_3$ edge in the energy dependence of the scattered x-rays at $T$=20\,K [Fig.~\ref{Fig2}d], indicating that the scattering is coming mainly from the Ta ions. We conclude that it arises from disordered local charge quadrupoles, which also give rise to scattering processes that preserve the x-ray polarization. The sharp drop in the peak intensity below $T_\mathrm{Q}$ arises from the splitting of the $(707)_t$ peak below $T_\mathrm{Q}$, as described in the earlier section.

To assist in interpreting the origin of the extra intensity of the $\textbf{Q}$=$(777)_\mathrm{c}$ reflection above $T_\mathrm{Q}$, we now perform DFT calculations with \ctc\, constrained to its cubic crystal structure.
We construct a supercell with disordered Ta local moments, as in Fig.~\ref{Fig2}b, but using the high temperature $Fm\bar{3}m$ experimental structure. Fig.\,\ref{Fig2}e shows the calculated components of the charge quadrupoles on the Ta sites for this case.
Note that while all the charge quadrupoles are inconsistent with the cubic crystal structure symmetry, they are non-zero because the spin lowers the local symmetry. 
We find that the variation of the charge quadrupoles in the cubic crystal structure is large, of the order of around 0.5 electron charges. Such large variations indicate that all charge quadrupole components can be finite and significant following the spin direction in the cubic environment due to the spin-orbit-entangled nature of the 5$d^1$ electron in \ctc.
Our analysis confirms that charge quadrupoles exist even in the cubic setting at $T>T_\mathrm{Q}$, consistent with our interpretation of the REXS data (Fig.\,\ref{Fig2}c). In our DFT model, these variations are the consequence of the spin moment fluctuations, however they would also be induced by local crystal structure distortions.

\begin{figure}[t!]
\includegraphics[width=3in]{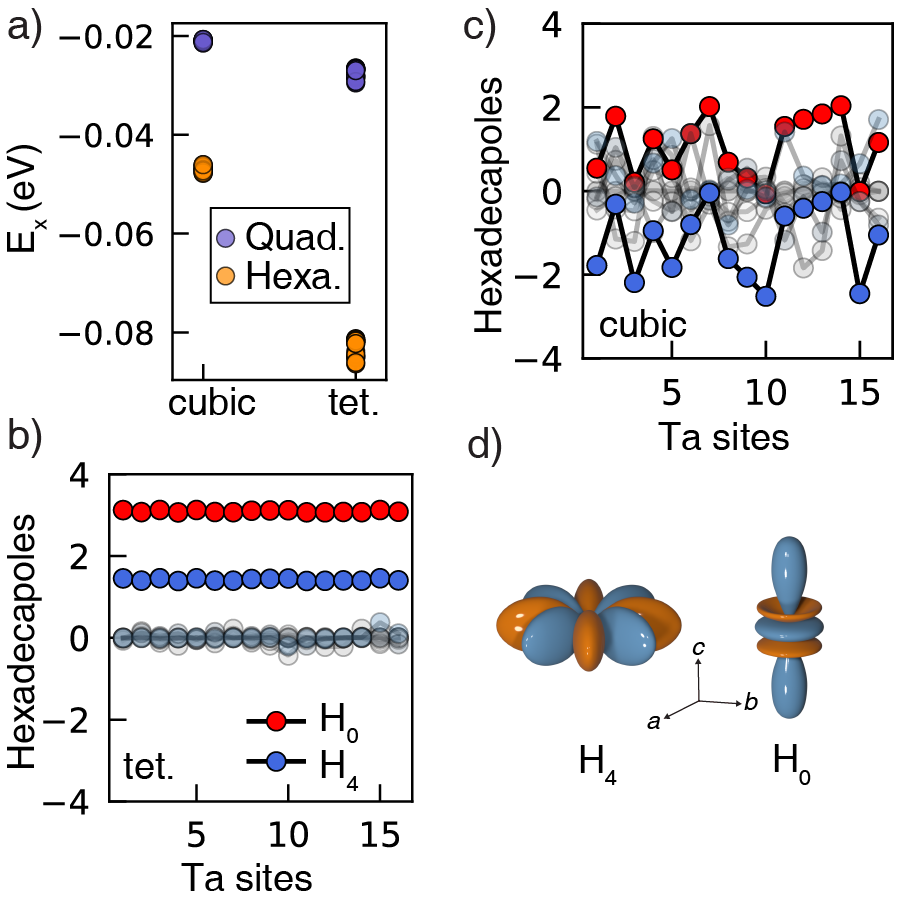}
\centering
\caption{a) Exchange energies of the charge quadrupoles and hexadecapoles calculated for cubic and tetragonal phases of \ctc. b) The calculated charge hexadecapoles in the tetragonal phase of \ctc\, for the atoms labeled in Figure 2a. c) The calculated charge hexadecapoles in the cubic phase of \ctc\, for the Ta atoms labeled in Figure 2a. d) The shapes of charge hexadecapole components, the $H_0$ and $H_4$.}
\label{Fig3}
\end{figure}

Finally, we follow the approach of Ref.~\onlinecite{bultmark_multipole_2009}, and calculate the contribution of the quadrupoles as well as the symmetry-allowed next-order charge multipoles, the hexadecapoles, to the total on-site exchange energy ($E_x$) for both cubic and tetragonal phases. Ref.~\onlinecite{bultmark_multipole_2009} demonstrated that the multipole channel corresponding to the most negative $E_x$ has the main energetic contribution to the phase transition and can be considered as a primary order parameter. Fig.\,\ref{Fig3}a shows the calculated $E_x$s on all 16 Ta ions (in the supercells) in both the cubic and tetragonal settings. Interestingly, in both crystal structures, charge hexadecapoles comprise the main contribution to the $E_x$ (the magnetic multipoles are not shown here as they do not order). Furthermore, we observe that the $E_x$ values for both charge quadrupoles and hexadecapoles are more negative for the tetragonal phase, indicating that the structural phase transition at $T_\mathrm{Q}$ stabilizes them further.

Here, we use a simplified notation for charge hexadecapoles (rank 4 tensor), $H_t$, where $-4\leqslant t \leqslant 4$ as described in the Ref.~\onlinecite{bultmark_multipole_2009} formalism. In Fig.\,\ref{Fig3}b,c we plot the charge hexadecapole components for the 16 Ta sites of the supercell for both the tetragonal and cubic symmetries of \ctc. When the symmetry of the supercell is constrained to the tetragonal symmetry (corresponding to $T < T_\mathrm{Q}$), two non-zero components of the charge hexadecapoles, $H_0$ and $H_4$ (illustrated in Fig.\,\ref{Fig3}d) order ferroically while all other components are calculated to be close to zero. Unlike the charge quadrupoles, however, the $H_0$ and $H_4$ components of the charge hexadecapoles tend to order ferroically in the cubic setting (corresponding to $T > T_\mathrm{Q}$) as shown in Fig.\,\ref{Fig3}c. Therefore, despite the large negative $E_x$, the charge hexadecapoles cannot be viewed as the primary order parameter associated with the transition.

\section{Summary}
In summary, we provide a comprehensive description of the coupled spin, charge, multipole, and structural order in \ctc. First, we synthesized a single crystal of \ctc\, and detected a new anomaly in the heat capacity measurements at temperature $T_\mathrm{Q}$ previously associated with a poorly understood phase transition. Then, we used single-crystal x-ray diffraction to determine the cubic-tetragonal structural changes at this temperature. Our DFT calculations confirmed that charge quadrupoles order ferroically for the tetragonal low-temperature phase, consistent with our measured compression of Ta--Cl bonds along z. In the region, $T > T_\mathrm{Q}$ our temperature-dependent REXS measurements showed a steady build-up of short-ranged order on approaching $T_\mathrm{Q}$. Our DFT calculations further corroborated the REXS observations suggesting that this could result from disordered local charge quadrupole moments in the cubic high-temperature phase driven either by local structural distortions or the paramagnetic nature of the spin moments at $T > T_\mathrm{m}$, or a combination of both.

In further work, a decisive proof of multipolar order with REXS, will require a material in which the structural peaks do not coincide with the resonant charge quadrupolar or magnetic dipolar peaks. In double perovskite $5d^1$ systems, this could be achieved if charge/magnetic multipoles order antiferroically. In addition, a large magnitude of multipole moments is desirable for successfully detecting hidden orders. Hence, an essential task in the subsequent investigations of hidden order in 5d$^1$ double perovskites that are amenable to the existing techniques is identifying new materials possessing large multipole moments with antiferroic 
long-ranged order.

\section{Methods}
\subsection{Computational details}
All DFT calculations were performed using the Vienna $ab-initio$ simulation package (VASP) using a plane-wave basis set. Projector augmented wave pseudopotentials\cite{kresse_ab_1993, kresse_efficient_1996, kresse_ultrasoft_1999, blochl_projector_1994}, Cs\_sv, Ta, and Cl as provided by VASP were used in conjunction with the Perdew-Burke-Ernzerhof (PBE) functional\cite{perdew_generalized_1996}. and on-site effective Hubbard $U_{eff}$ = 2\,eV correction for the Ta $5d$ orbitals\cite{dudarev_electron-energy-loss_1998}. Spin-orbit coupling was included explicitly. An energy cutoff of 600 eV and gamma-centered k-point meshes of 8$\times$8$\times$6 and 2$\times$2$\times$1 were used for the tetragonal unit cell and the supercells, respectively.
The multipole moments were calculated by decomposing the DFT charge \cite{cricchio_itinerant_2009} or spin \cite{spaldin_monopole-based_2013} densities in the spheres around the atomic sites into their component charge or magnetic multipoles. In this implementation, the local multipole moments are obtained by decomposing the DFT density matrix into the components of the irreducible spherical tensor centered on the ions. For details of the formalism, see Ref.~\onlinecite{cricchio_itinerant_2009}. The local on-site exchange energy, $E_x$, is reformulated in terms of the multipolar decomposition of the density matrix following Ref.~\onlinecite{bultmark_multipole_2009}.
This method has already been implemented in the all-electron full-potential linearised augmented-plane wave ELK code\cite{elk}. Here, we have implemented this method within an open-source python code compatible with any electronic structure calculation code.

\subsection{Synthesis and Characterization}

Polycrystalline \ctc\, was synthesized by the solid-state reaction method described in Ref.~\onlinecite{ishikawa_ordering_2019}. To identify structural changes in \ctc, synchrotron x-ray diffraction was performed on polycrystalline \ctc\, at the MS-powder beamline at the Swiss Light Source (PSI). The powder was filled in borosilicate glass capillaries with a diameter of 0.3 mm, which were then closed with grease and sealed before the experiment. Measurements were performed with an incident x-ray wavelength of 0.563\AA\, and to temperatures down to 4.2 K (See Fig.S1). The heat capacity measurements of poycrystalline \ctc\, was performed on a Physical Properties Measurements System (PPMS, Quantum Design) down to $T$=2\,K, as shown in Fig.S2.

The Cs$_{2}$TaCl$_{6}$ single crystals were synthesized following the procedure for K$_{2}$TaCl$_{6}$ in Ref.~\onlinecite{ishikawa_ordering_2019}. CsCl and TaCl$_{5}$ powders were mixed in a ratio of 2:1 and placed in a quartz tube with Ta metal wire. The tube was evacuated using a mechanical and turbo pump, then sealed and heated to 600 $^{\circ}$C, where it was held for 36 hours. The system was then cooled down to 500 $^{\circ}$C with a slow cooling rate of 0.5 $^{\circ}$C/h. Since the starting materials and final product are air-sensitive, all processes were performed in the glove box with Ar atmosphere. Dark purple single-crystal plaquettes were obtained and could be cleaved easily along the (111) plane. The quality of the single crystals was checked using a Rigaku Synergy-I XtaLAB x-ray diffractometer. Heat capacity measurements on the \ctc\, single crystals were also performed on a PPMS down to $T$=2\,K.

To look for evidence of short-ranged charge quadrupolar order, resonant elastic x-ray scattering (REXS) was performed on a single crystalline \ctc\, on the 6-circle diffractometer in the EH1 hutch of the P09 hard x-ray beamline at the PETRA III storage ring (DESY). The incident x-ray energy was tuned to close to the Ta $L_3$ absorption edge ($E\sim9881$ eV), and the polarization of the scattered beam (in vertical scattering geometry) was analyzed with a Cu(400) analyzer to differentiate between the $\sigma$$\to$$\sigma^\prime$ and $\sigma$$\to$$\pi^\prime$ scattering events. As \ctc\, is moderately air-sensitive, the single crystal was freshly cleaved just prior to the REXS measurement.

\section{Acknowledgments}
This work was funded by the European Research Council (ERC) under the European Union’s Horizon 2020 research and innovation program projects HERO (Grant No. 810451) and CALIPSOPlus (Grant No. 730872). The beamline proposal numbers for the \ctc\, experiments are I-20210904 EC (P09, DESY) and 20210818 (MS-POWDER, SLS). We thank S. Francoual and A. Cervellino for assistance during the various experiments. Calculations were performed at the Swiss National Supercomputing Centre (CSCS) under project IDs s889 and eth3 and on the EULER cluster of ETH Zürich. JRS acknowledges support from the Singapore National Science Scholarship from the Agency for Science Technology and Research.

\bibliography{Aria2, Ref, Max, journals2, books}
\end{document}